\documentclass[twocolumn,prl,superscriptaddress,aps,longbibliography,appendices]{revtex4-1}
\usepackage{amsmath}
\usepackage{mathtools}
\usepackage{physics}
\usepackage{graphicx}
\usepackage[per-mode=symbol]{siunitx}
\usepackage{ulem}
\DeclareSIUnit\gauss{G}
\DeclareSIUnit\bohr{a_{B}}
\DeclareSIUnit\ctms{ct/ms}
\DeclareSIUnit\cps{cps}

\usepackage{hyperref}
\usepackage{bbm}
\usepackage{soul}

\usepackage{color}
\definecolor{mygreen}{rgb}{0,0.5,0} 
\definecolor{mygrey}{rgb}{0.5,0.5,0.5} 
\definecolor{myred}{rgb}{0.75,0,0} 
\definecolor{myblue}{rgb}{0,0,0.75} 
\definecolor{mymagenta}{cmyk}{0,1,0,0.12} 
\definecolor{mycyan}{cmyk}{1,0,0,0.12} 
\definecolor{myorange}{rgb}{0.85,0.375,0}  
\definecolor{myviolet}{rgb}{0.5,0.3,1} 
\definecolor{mybrown}{rgb}{0.542969,0.269531, 0.0742188} 

\usepackage{ulem}

\newcommand{\gtext}[1]{{\color{mygreen}#1}}

\newcommand{\osout}{\bgroup\markoverwith
{\textcolor{myorange}{\rule[0.5ex]{2pt}{0.4pt}}}\ULon}

\newcommand{\etanoise}{\eta_\mathrm{noise}}
\newcommand{\pathphase}{\phi_\mathrm{path}}
\newcommand{\atomphase}{\phi_\mathrm{cm}}
\newcommand{\Ndet}{N_\mathrm{det}}

\newcommand{\Ncoll}{N_\mathrm{coll}}
\newcommand{\Ncollbar}{\bar{N}_\mathrm{coll}}
\newcommand{\Npr}{N_\mathrm{pr}}
\newcommand{\Nprbar}{\bar{N}_\mathrm{pr}}
\newcommand{\Nprbarbar}{\bar{\bar{N}}_\mathrm{pr}}
\newcommand{\Nsc}{N_\mathrm{sc}}
\newcommand{\Nscbar}{\bar{N}_\mathrm{sc}}
\newcommand{\Nscbarbar}{\bar{\bar{N}}_\mathrm{sc}}
\newcommand{\Vbar}{\bar{V}}
\newcommand{\Vbarbar}{\bar{\bar{V}}}
\newcommand{\Cbar}{\bar{C}}
\newcommand{\Cbarbar}{\bar{\bar{C}}}
\newcommand{\temperature}{\mathcal{T}}
\newcommand{\That}{\hat{\temperature}}
\newcommand{\Fnoise}{F_{n}}
\newcommand{\EndMatter}{the End Matter}

\begin{document}
\include{./NewCommands}
\newcommand{\mytitle}{A single atom emitting resonance fluorescence into a coherent beam, and its use for non-destructive atom thermometry}

\title{\mytitle}

\newcommand{\ICFO}{ICFO - Institut de Ciencies Fotoniques, The Barcelona Institute of Science and Technology, 08860
Castelldefels (Barcelona), Spain}
\newcommand{\ICREA}{ICREA - Instituci\'{o} Catalana de Recerca i Estudis Avan{\c{c}}ats, 08010 Barcelona, Spain}

\author{Tom\'{a}\v{s} Lamich}
\email{tomas.lamich@icfo.eu}\email{lamichtom@gmail.com} 
\affiliation{\ICFO}
\author{Laura Zarraoa}
\affiliation{\ICFO}
\author{Sondos Elsehimy}
\affiliation{\ICFO}
\author{Morgan W. Mitchell}
\email{morgan.mitchell@icfo.eu}
\affiliation{\ICFO}
\affiliation{\ICREA}
\author{Romain Veyron}
\email{romain.veyron@icfo.eu}
\affiliation{\ICFO}

\begin{abstract}

Using a far-off-resonance optical dipole trap, we place a single neutral $^{87}$Rb atom in a weak, atom-resonant coherent beam, while also strongly illuminating it from an orthogonal direction to produce resonance fluorescence. The atom-modified coherent beam is then collected and its photon statistics analyzed. We observe first-order interference that can increase or decrease the beam flux, depending on the relative phase of the coherent beam and resonance fluorescence. This confirms predictions of Goncalves et al. [Phys. Rev. A 104, 013724]. The interference visibility is also shown to be a sensitive, time-resolved, non-destructive thermometer: by fitting the resulting photon count distributions, we infer the center-of-mass localization of the atom within the trap.  With 1200 atoms and integration time of \SI{80}{\ms} per atom, we demonstrate  temperature uncertainties of \SI{4}{\percent} for $\sim \SI{30}{\micro\kelvin}$ temperatures at $\sim\SI{200}{\micro\second}$ time resolution. 
\end{abstract}

\maketitle

In recent decades, individual trapped atoms, both charged \cite{Foss-FeigAR2025} and neutral \cite{HenrietQ2020}, have become an important platform for quantum computation and simulation, with potential application also in quantum networking \cite{CoveyNPJQI2023}. They are also important model systems for studying light-matter interactions with individual-particle resolution \cite{AljunidPRL2013, VolzLP2007, RosenfeldPRL2017, HoferPRXQuantum2025j}. In all these applications, center-of-mass temperature is an important parameter that affects for example, the fidelity of quantum gates \cite{Azuma_2024, Han2025} and storage times in quantum memories \cite{Cho2016}.
Measuring the temperature of a single atom is challenging, due to very weak signals and the intrinsically statistical nature of the measurement. Destructive thermometry methods, in which the atom can be lost and another must be substituted, include release and recapture (R\&R)  \cite{Tuchendler2008} and  time-of-flight (TOF)  \cite{Chu1985,Lett1988}. Non-destructive methods include scattering during resolved sideband cooling \cite{Monroe1995,Kaufman2012,Spence_2022} and self-interference of resonance fluorescence light \cite{Eschner2001, Slodicka2012}. 

Here we study light-matter interaction scenario first analyzed by Goncalves et al. \cite{Goncalves2021}: a two-level atom is resonantly excited by a ``pump'' beam from one direction, thus creating a coherently-radiating single quantum emitter, capable of both giving energy to and taking energy from the field. This radiator is positioned within a coherent ``probe'' beam from a different direction but at the same frequency. This scenario mixes dynamical effects (absorption and stimulated emission of the probe) with interference between the various processes that can put light into the probe mode. Here we study the statistical properties of this interference. Building upon \cite{Goncalves2021}, we describe the statistics of $G^{(1)}$ photon counting signals in the outgoing probe light, in the presence of (fast) atomic motion within the trap and (slow) drifts of path lengths and other experimental parameters. We find that the first-order correlations can serve as a non-destructive, time-resolved thermometer for the atomic center-of-mass motion, which determines the visibility as other parameters vary.


\begin{figure}[t]
    \centering
    \includegraphics[width=1\linewidth]{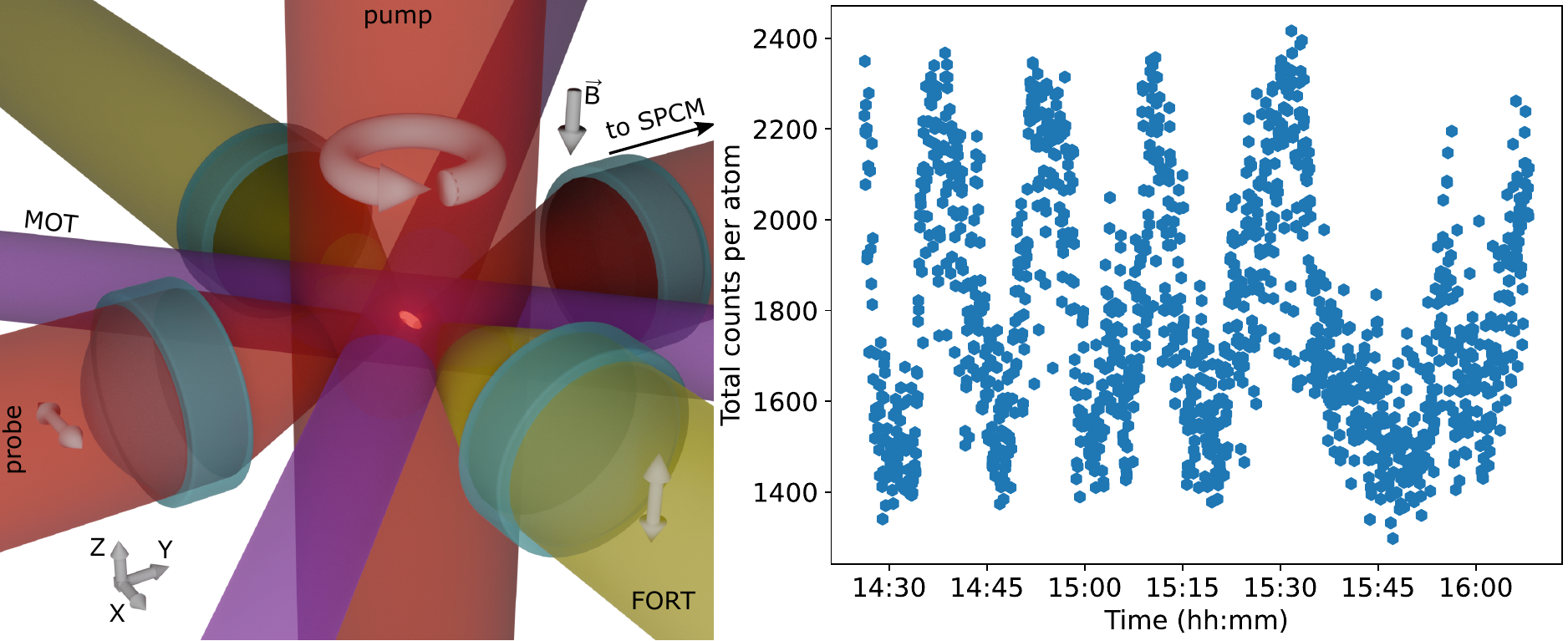}

    \caption{Left: Experimental geometry. Four high numerical-aperture lenses surround a small magneto-optic trap (MOT, purple, vertical beams not shown) which loads single atoms into a far-off-resonance trap (FORT, yellow). Pump and probe beams (red) illuminate the atom. Probe light, together with resonance fluorescence, are collected into single-mode fiber (not shown) and detected with a single photon counting module (SPCM). Polarizations are indicated by white arrows. A magnetic field is applied along $-Z$ to support Zeeman optical pumping. The single-atom centre-of-mass distribution after cooling is indicated as the bright, cigar-shaped region (not to scale) at center. Right: collected photons show interference as path lengths slowly drift. Each point shows the total photons detected from a single atom in $80\times \SI{1}{\milli\second}$ illumination intervals, interspersed with cooling.  
    }
    \label{fig:GeometryAndSinglesData}
\end{figure}

\begin{figure}[t]
    \centering
\includegraphics[width=1\linewidth]{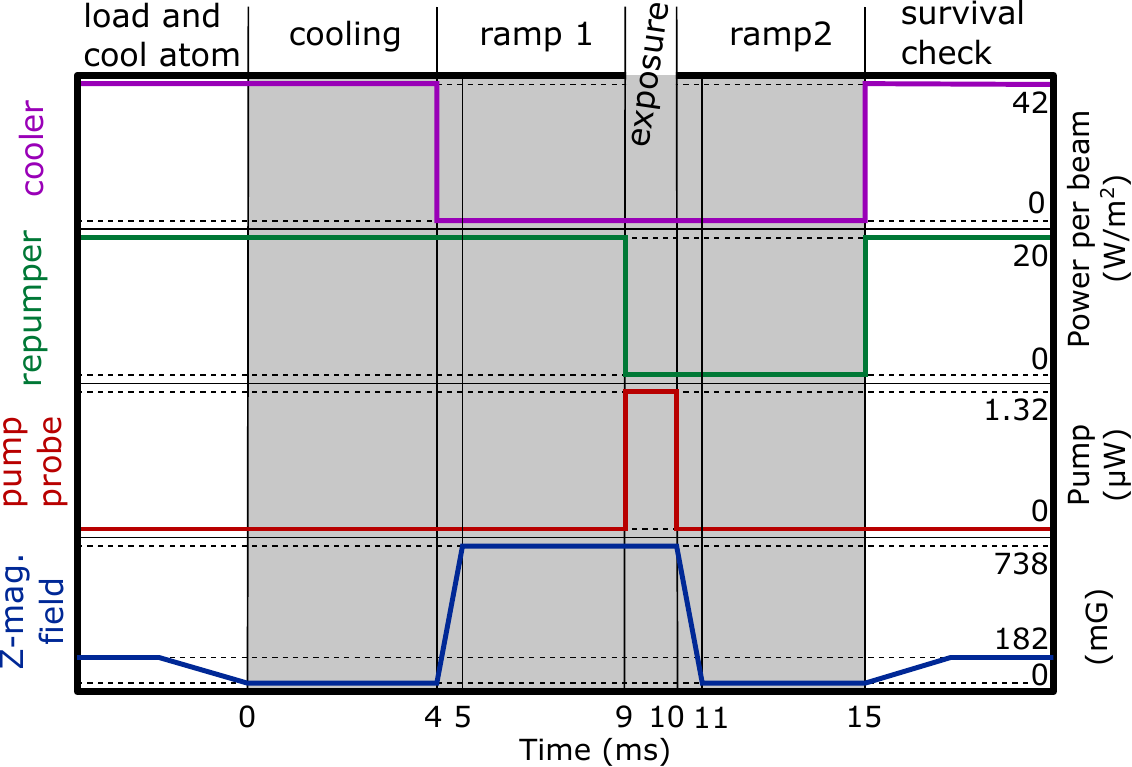}
\caption{
Experimental sequence. An atom is loaded to the trap and cooled to the minimum achievable temperature using PGC. The magnetic field is then ramped up and the atom is pumped to $F=2, m_F=-2$. During exposure the atom is illuminated with pump and probe beams and the arrival times of photons in the probe mode are recorded. The magnetic field is then ramped back to zero. The gray shaded area is repeated $80$ times. Finally, the MOT beams are turned on and the resulting fluorescence is observed to verify the atom remained trapped during the sequence.
    }
    \label{fig:Sequence}
\end{figure}

As illustrated in  \autoref{fig:GeometryAndSinglesData} (left), we use a magneto-optical trap (MOT) to load single $^{87}$Rb atoms into a strongly-focused far-off-resonance trap (FORT), as described in \cite{Bruno2019, Bianchet2022ORE}, with additional detail in \EndMatter. In the FORT, the atom experiences polarization gradient cooling (PGC) \cite{KaufmanPRX2012,JoshiIOP2020,ClementsPRL2026} before being optically pumped as described in \EndMatter, and then \gtext{it is} simultaneously illuminated with a strong pump beam and a weak probe. The $Z$-propagating pump is circularly-polarized and resonant to the closed $F, m_F = 2,-2 \to F, m_F = 3,-3$ transition, so the atom can repeatedly scatter photons, some of which enter the $Y$-propagating, $X$-polarized probe beam, which is strongly focused on the trap center and thus on the position of the atom. The pump and probe derive from the same laser and share the same frequency. The scattering geometry ensures that the probe and scattered light have the same polarization. Collection with a high numerical-aperture (NA) lens into single-mode fiber ensures spatial overlap. The collected photons are detected with a single-photon counting module (SPCM) and time-tagged with a field-programmable gate array (FPGA). As shown in \autoref{fig:Sequence}, we load an atom then alternate PGC with \SI{1}{\ms} exposure to pump and probe for {80} cycles, counting the resulting probe photons. This shows oscillation with the pump-probe relative phase at the atom, as shown in \autoref{fig:GeometryAndSinglesData} (right).

\begin{figure}[!t]
    \centering
    \includegraphics[width=0.85\columnwidth]{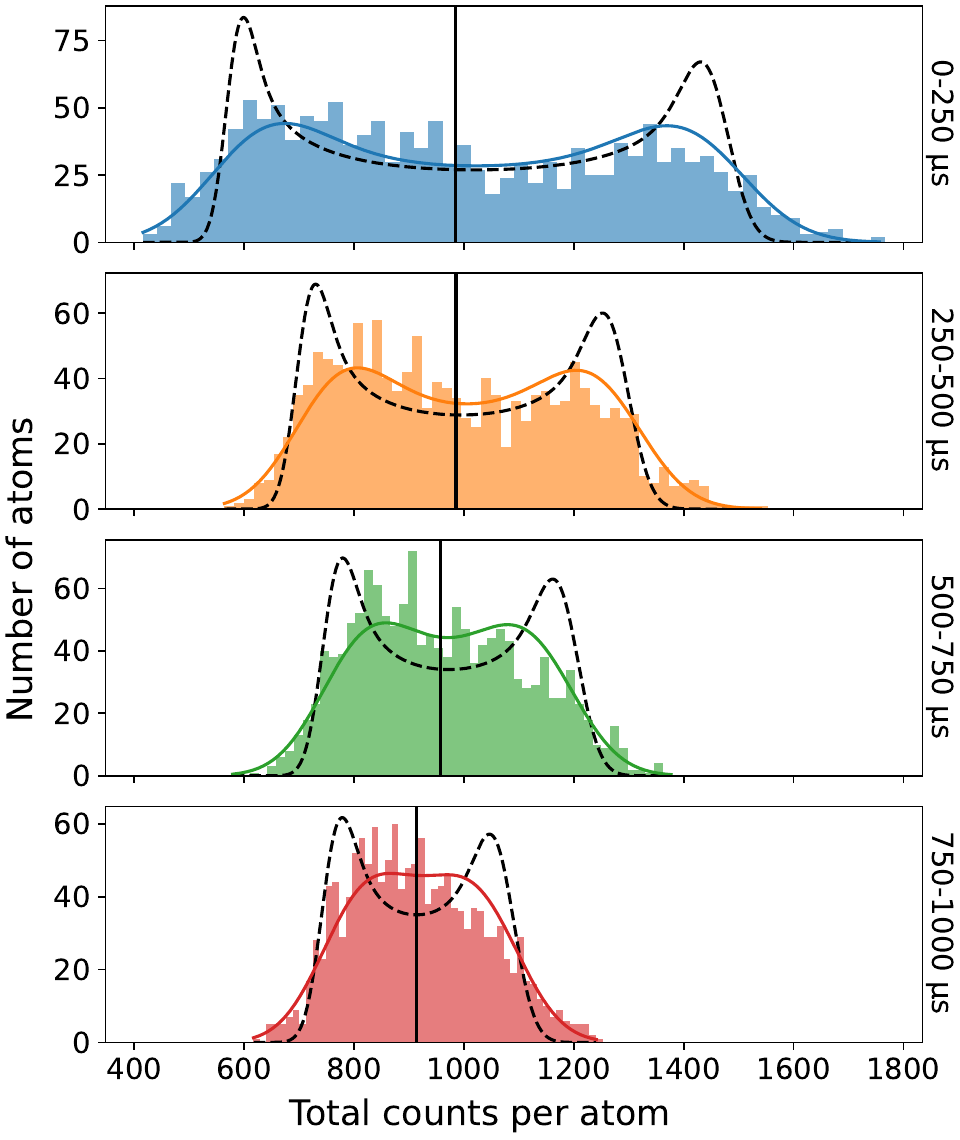}
    \caption{Histograms of detected counts in \SI{250}{\us} time bins (indicated at right) of a \SI{1}{\ms} exposure window, acquired as 80 windows per atom for 1200 atoms over \SI{1.5}{\hour}. Black vertical lines show the mean number of collected counts. Curves show fits with \autoref{eq:AP_dist_integral_maintext}, dashed curves with $\sigma_n = 0$, i.e., assuming no fluctuations in experimental conditions.  The fitted curves are scaled to have mean probability density equal to that of the data histograms.  Each histogram is plotted with $\approx 50$ bins so that the bin edges fall on integers.
    }
    \label{fig:hist_sequence}
\end{figure}

The interference visibility in the exposure time window depends strongly on the localization of the atom, and thus provides information on the atom temperature. This is illustrated in  \autoref{fig:hist_sequence}, which shows histograms of photon counts during the four quartiles of a \SI{1}{\milli\second} illumination. As above, we collect atoms, cool them, and then expose them to pump and probe for \SI{1}{\milli\second}, time-tagging the SPCM detections, with PGC re-cooling between each exposure interval. We group the detections into \SI{250}{\micro\second} time bins and sum the counts produced by a given atom in each time bin. Repeating for many atoms we generate a histogram. The results clearly show a reduction in interference visibility for later time bins, consistent with an increase in temperature during the exposure.  

\begin{figure}[t]
    \centering    \includegraphics[width=0.85\columnwidth]{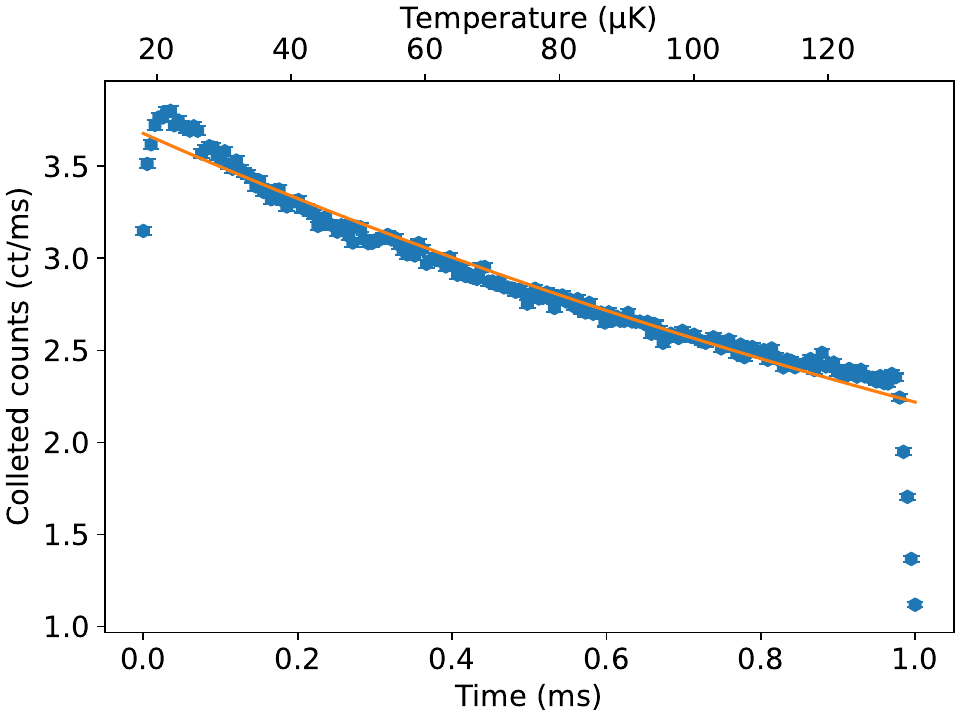}
    \caption{Average scattering rate from a single atom induced by the pump beam. The scattering rate exhibits nearly linear decay in time due to increase of atom temperature. The bottom axis shows the elapsed time, while the top axis shows the estimated atom temperature. The error bars show 1 standard error of the mean. The solid orange line shows the fitted model given by \autoref{eq:scat_decay_exp}.}
    \label{fig:temp_mapping}
\end{figure}

R\&R thermometry, implemented as in \cite{Bianchet2022ORE}, finds the atom temperature to be $\temperature_\mathrm{init} \equiv \SI{18(2)}{\micro\K}$ before exposure (i.e., after PGC cooling) and $\temperature_\mathrm{end} \equiv \SI{133(14)}{\micro\K}$ after \SI{1}{\milli\second} of pump-probe exposure. To interpolate these, we measure the pump-only scattering rate $\Nsc(t)$ versus time $t$, fit with an exponential (see \EndMatter~\autoref{fig:temp_mapping}), and compute the resulting temperature as $\temperature= \temperature_\mathrm{init} + \alpha \int_0^t \Nsc(t') dt'$, where $\alpha$ is chosen such that $\temperature = \temperature_\mathrm{end}$ at $t=\SI{1}{\ms}$.

As the atom heats, we observe two effects: first, there is a reduction of the interference visibility as the atom explores a wider area in the FORT. The center of the distribution also shifts towards fewer counts, which we attribute to reduced light shifts \cite{Chin2017} and reduced collection efficiency as the atom is on average farther from the lens focus. 

To model the interference behavior, we extend the model of Goncalves et al. \cite{Goncalves2021}, to include  photon count distributions under fluctuating experimental conditions. We model the trapped atom as a two-level system with ground state $\ket{g}$ and excited state $\ket{e}$ driven by a resonant field with a Rabi frequency $\Omega=\Omega_\mathrm{pump}e^{i\phi}+\Omega_\mathrm{pr}$, where $\Omega_\mathrm{pump}$ and $\Omega_\mathrm{pr}$ are the Rabi frequencies induced by the coherent pump and probe fields, respectively, and $\phi$ is the phase difference between the two fields {at the atom}.  From the first order correlation function of the total electric field in the detection mode, the detected photon flux per unit of time $\Phi_\mathrm{det}$ can be written as
\begin{equation}
    \label{eq:N_det_def}
    \Phi_\mathrm{det}=\Phi_\mathrm{pr}+\Gamma_0\rho_{ee}-2\sqrt{\Gamma_0 \Phi_\mathrm{pr}} \mathrm{Im}(\rho_{eg}),
\end{equation}
where $\Phi_\mathrm{pr}$ is the  probe flux and $\Gamma_0$ is the natural decay rate. In the low saturation regime, the density matrix elements at steady state are $\rho_{ee}=|\Omega|^2/\Gamma_0$ for the excited state population and $\rho_{eg}=i\Omega/\Gamma_0$ for the optical coherence. 

Denoting with $\Ncoll$ the mean number of collected photons over a duration $\Delta{t}$, from \autoref{eq:N_det_def}, in the limit where the scattering is dominated by the pump beam, $|\Omega_\mathrm{pr}|\ll|\Omega_\mathrm{pump}|$, we find that 
\begin{equation}
\label{eq:detcted_no_photons}
\Ncoll=N_\mathrm{sc}+N_\mathrm{pr}-2\sqrt{N_\mathrm{sc}N_\mathrm{pr}}\cos \phi,
\end{equation}
where the collected atom scattering from pump alone is $N_\mathrm{sc}=\eta \Delta{t} |\Omega_\mathrm{pump}|^2/\Gamma_0$ and the number of detected probe photons is $N_\mathrm{pr}=\eta \Phi_\mathrm{pr}\Delta{t}$, and $\eta$ is the detection efficiency. 

The phase $\phi$ is given by the wave-number $2\pi/\lambda$, where $\lambda$ is the pump and probe wavelength, times the differential length between two paths: from the laser to the detector via the probe, and from the laser to the detector by scattering at the position of the atom. If pump and probe are at right angles, $\phi = \pathphase + \atomphase$, where $\pathphase$ is the path difference to the center of the trap and $\atomphase = 2 \pi (d_Y-d_Z) / \lambda$, where $d_Y$ and $d_Z$ are the atomic center-of-mass (cm) displacement along the pump and probe propagation directions, respectively.  Assuming a Boltzmann distribution in a $Y-Z$ symmetric harmonic trap, $\atomphase$ is normally distributed with variance $\sigma^2_\phi = 2 k_B\temperature/(m\omega^2)$, where $\temperature$ is the atom temperature, $k_B$ is the Boltzmann constant, $m$ is the atomic mass, and $\omega$ is the trap frequency.

As seen in \autoref{fig:GeometryAndSinglesData} (right), $\pathphase$ varies over $\sim \SI{}{\minute}$ time scales due to temperature drifts, whereas $\atomphase$ varies over $\sim\SI{}{\micro\second}$ time scales as the atom moves in the trap. We consider the photon counting statistics for repeated acquisitions of $\sim \SI{}{\milli\second}$ duration, such that $\atomphase$ must be averaged over.  We will use a single overbar to indicate averages over times long enough for the atom to explore the trap's available phase space, and a double overbar to indicate averages over longer times. Averaging \autoref{eq:detcted_no_photons} over $\phi_\mathrm{cm}$, the mean collected photons in a \SI{1}{\ms} window is 
\begin{equation}    \label{eq:number_of_detected_phtons_averaging}
    \begin{aligned}
\Ncollbar=& \Nscbar +\Nprbar - 2\sqrt{\Nscbar \Nprbar} \Cbar \cos \pathphase,
    \end{aligned}
\end{equation}
with $\Cbar \equiv \exp [-\sigma_\phi^2/2]$. This describes interference with visibility
\begin{equation}
V \equiv \frac{b-a}{b+a}  = \frac{2\sqrt{\Nscbar \Nprbar}}{\Nscbar + \Nprbar} 
    \Cbar, 
\end{equation}
where $a \equiv \min_\phi{\Ncollbar}$ and $b \equiv \max_\phi{\Ncollbar}$ are the lower and upper bounds on $\Ncollbar$.

We now consider the statistics of $\Ncollbar$ over many such \SI{1}{\milli\second} acquisitions, spread over time such that $\Nscbar$, $\Nprbar$, $\Cbar$ and $\phi_\mathrm{path}$ can vary. As described in \EndMatter, if we allow for fluctuations of  $\Nscbar$, $\Nprbar$ and $\Cbar$ about their long-time means $\Nscbarbar$, $\Nprbarbar$ and $\Cbarbar$, respectively, the net effect is, to lowest order, an additive noise $\etanoise$ such that 
\begin{equation}    \label{eq:NcollbarInclNoise}
    \begin{aligned}
\Ncollbar = &\Nscbarbar +\Nprbarbar - 2\sqrt{\Nscbarbar \Nprbarbar} \Cbarbar \cos \pathphase +\etanoise,
    \end{aligned}
\end{equation}

Finally, we consider the photon-counting statistics: For a given $\Ncollbar$, the number of detected photons $\Ndet$ follows a Poisson distribution $P_{\Ncollbar}(\Ndet)$, resulting in a distribution for $\Ndet$, the detected number of photons: 
\begin{equation}
\label{eq:AP_dist_integral_maintext}
    P(\Ndet) = \int P_{\Ncollbar}(\Ndet) (F_\mathrm{a,b} * \Fnoise) (\Ncollbar)d\Ncollbar, 
\end{equation}
where $F_\mathrm{a,b} * \Fnoise$ is the convolution of the noise probability density function (PDF) $\Fnoise$ with the arcsine PDF
\begin{equation}
   \label{eq:arcsine}
   F_\mathrm{a,b}(\Ncollbar) = \left\{ 
\begin{array}{cc}
\frac{1}{\pi\sqrt{(\Ncollbar-a)(b-\Ncollbar)}}  &  a < \Ncollbar < b \\\ 0 & \mathrm{otherwise} 
\end{array} \right.
\end{equation}
with limits $a = \Nscbarbar + \Nprbarbar - 2\sqrt{\Nscbarbar \Nprbarbar} \Cbarbar$,  $b = \Nscbarbar + \Nprbarbar + 2\sqrt{\Nscbarbar \Nprbarbar} \Cbarbar$. In practice, we take $\etanoise$ to be zero-mean, normally distributed with variance $\sigma^2_n$. 
Detector dark counts are not included; in the experiment their rate is negligible.

We use independent measurements of $\Nprbarbar$ and $\Nscbarbar$, by illuminating with only pump or only probe, respectively, to find the pre-factor $2\sqrt{\Nscbarbar\Nprbarbar}/(\Nscbarbar+\Nprbarbar)$. We then illuminate with both pump and probe and collect statistics to form histograms of $\Ndet$. These are maximum-likelihood fitted with  \autoref{eq:AP_dist_integral_maintext} to find  $\sigma_n^2$, $a$ and $b$. We then calculate the visibility $\Vbarbar \equiv ({b-a})/({b+a})$ and the ``visibility temperature'' estimator
\begin{equation}
    \label{eq:temp_on_contrast}
     \That \equiv \frac{\lambda^2}{4\pi^2}\frac{m\omega^2}{k_B}\ln \frac{2\sqrt{\Nscbarbar \Nprbarbar}}{\Vbarbar(\Nscbarbar + \Nprbarbar)}.
\end{equation}

\begin{figure}
    \centering    
    \includegraphics[width=0.9\linewidth]{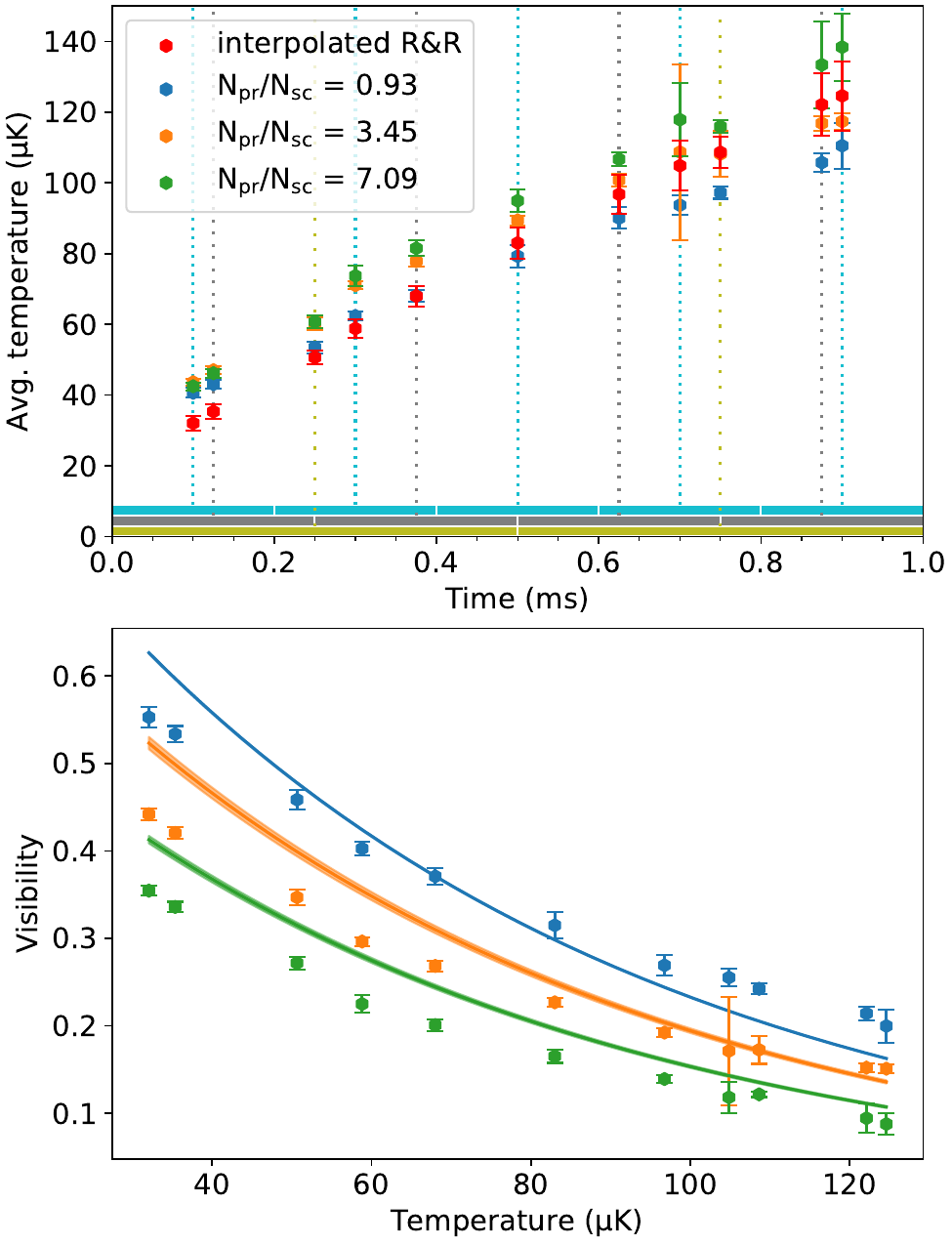}
    \caption{Thermometry from interference visibility, estimated from photon-count histograms acquired as for \autoref{fig:hist_sequence}. 
    (upper) $\That$ from Eq. \eqref{eq:temp_on_contrast}, applied to best-fit parameters versus time within a time bin, for three different ratios of probe and scattered-light brightness. Colored horizontal bars show time bins with dotted vertical lines indicating their centers. The $\Npr/\Nsc$ ratios are calculated using the average $\Npr$ and $\Nsc$ during the \SI{1}{\ms} exposure window.
    (lower) Visibility $\Vbar$ versus temperature, interpolated between R\&R measured values at \SI{0}{\milli\second} and \SI{1}{\milli\second} end-points. Solid curves show $\Vbar$. 
    Shaded areas around the curves show $\pm 1\sigma$ uncertainty on the prediction, due to uncertainties of $N_\mathrm{pr}$ and $N_\mathrm{sc}$. Points show the best-fit value for $\Vbar$ from histograms. Error bars show $\pm 1\sigma$ uncertainty from fit covariance matrices.
    }

\label{fig:contrasr_v_temperature}
\end{figure}

\begin{figure}
    \centering
    \includegraphics[width=0.85\columnwidth]{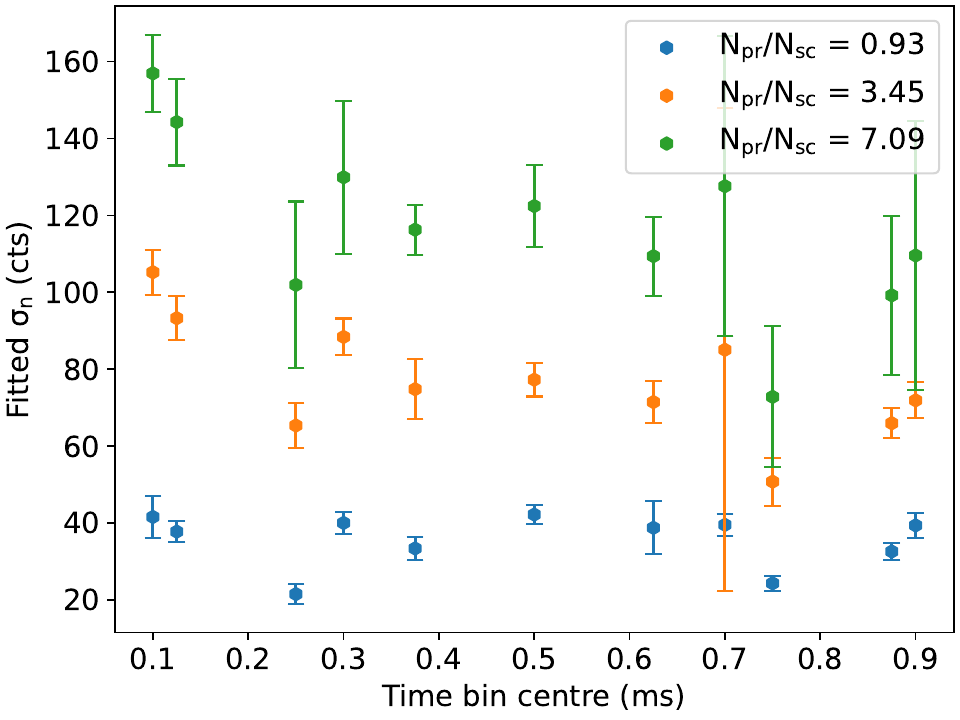}
    \caption{Fitted $\sigma_n$ of the Gaussian distribution, see \autoref{eq:AP_dist_integral_maintext}, versus the used time-bin centers. Error bars show $\pm 1\sigma$ uncertainty on the fitted parameter. }
    \label{fig:fitted_sigmas}
\end{figure}

The model predicts decreasing visibility with increasing atom temperature and shows good agreement with the established R\&R thermometry method. As for \autoref{fig:hist_sequence}, we split the exposure time into time bins of \qtylist{200;250;500}{\micro\second}, then the resulting histograms are fitted using \autoref{eq:AP_dist_integral_maintext}, and $\That$ is computed using \autoref{eq:temp_on_contrast}. In this way we measure the atom temperature throughout the illumination interval.  \autoref{fig:contrasr_v_temperature} (upper) shows a comparison of the temperature interpolated from R\&R measurements at \SI{0}{\ms} and at \SI{1}{\ms}, and this new method. The visibility temperature is robust with respect to the ratio $N_\mathrm{pr}/N_\mathrm{sc}$, and agrees well with the R\&R temperature. The technical broadening $\sigma_n$ does not depend strongly on temperature, suggesting that the additive noise is primarily from variation of the collected photon numbers $\Nprbar$ and $\Nscbar$, rather than of $\Cbar$, see \EndMatter,  \autoref{fig:fitted_sigmas}.

Here we used 1200 atoms, each with eighty \SI{1}{\ms} illumination periods. When grouped into  $\SI{200}{\us}$ bins, and thus giving $\SI{200}{\us}$ time resolution, we observe a bin-averaged relative uncertainty of $4\%$. For comparison, the R\&R thermometry gave $\approx 10\%$ relative uncertainty, while also destroying the atomic state. The time resolution is limited by the fitting process, which becomes unstable at low photon numbers and for atomic temperatures higher than $\sim\SI{70}{\micro\K}$. We note the interference signal could be used to stabilize paths to a single emitter. If that were done, a possible refinement of the technique would control the pump-probe phase $\pathphase$ to measure only with $\pathphase\in\{0,\pi\}$, making $\Ncollbar$ maximally sensitive to $\Cbarbar$ and thus to temperature.

We have demonstrated amplification and de-amplification of a coherent beam by a single-atom undergoing resonance fluorescence, and its use for non-destructive thermometry of a single trapped atom. The method could be implemented in many trapped neutral atom and ion systems, and could be used to optimize cooling methods such as feedback cooling  \cite{Koch_PRL_2010, Bushev_PRL_2006}.  A variant in which the probe is introduced after collection, for example by displacing the collected field with a weak beamsplitter and a coherent state, could make the method available to experiments in which the single emitter is only strongly coupled to one traveling mode of the field. Although here we operated in a high-visibility regime with similar fluxes of probe and scattered photons, in principle the method can also be used with a high flux of probe photons, at which point the process resembles homodyne detection of the resonance fluorescence field, of interest for studies of advanced quantum dynamics \cite{PinolPRR2024, settimo2026quantumjumpunravelingsnonmarkovian}. 


\begin{acknowledgements}
Acknowledgments: We thank D. Gon\c{c}alves and D. Chang for helpful discussions. Supported by European Commission projects Field-SEER (ERC 101097313) and QUANTIFY (101135931); Horizon Europe (Q-Planet, 101291743), Chips Joint Undertaking (Chips JU) and the Spanish Ministry for Digital Transformation and Civil Service under reference CJU-010200-2026-6, as part of the Recovery, Transformation and Resilience Plan (PRTR),  Spanish Ministry of Science MCIN projects SAPONARIA (PID2021-123813NB-I00) and SALVIA (PID2024-158479NB-I00),  ``Severo Ochoa'' Center of Excellence CEX2024-001490-S [MICIU/AEI/10.13039/501100011033];  Generalitat de Catalunya through the CERCA program,  DURSI grant No. 2021 SGR 01453 and QSENSE (GOV/51/2022).  Fundaci\'{o} Privada Cellex; Fundaci\'{o} Mir-Puig.
TL acknowledges Marie Sk\l{}odowska-Curie grant agreement No 847517.  Views and opinions expressed are those of the authors only and do not necessarily reflect those of the European Union or the European Research Council Executive Agency. 
\end{acknowledgements}

\textit{Data availability:} The data that support the findings of this study are available from the corresponding author upon reasonable request. 

\bibliographystyle{apsrev4-1no-url-title} \bibliography{thermometry.bib}

\clearpage
\onecolumngrid
\begin{center}
    \large\bfseries End matter for: \mytitle
\end{center}
\twocolumngrid

\newcommand{\PRLsection}[1]{\noindent {\it#1} -- }

\PRLsection{Experimental methods}
{The FORT beam is generated by an \SI{852}{\nm} diode laser, power-stabilized to give \SI{12}{\mW} before the vacuum chamber, polarized along $Z$, and focused by an NA = 0.5 aspheric lens to a \SI{1.32}{\um} waist, measured via quantum jump spectroscopy \cite{Bianchet2022QJS}. This keeps the transverse trap frequencies constant at $\omega=2\pi\times\SI{104}{\kHz}$.  The atom survival probability was $\approx\SI{75}{\%}$ after the total sequence duration of $\SI{1.305}{\s}$, consistent with the background-gas-limited lifetime. The pump, probe and the cooler light come from a single laser diode, stabilized on the $F=2\to F'=3$ transition of the D$_2$ line of the $^{87}$Rb atom using modulation transfer spectroscopy \cite{MartinezdeEscobar2015}. The repumper laser is offset-locked to the cooler near the $F=1\to F'=2$ D$_2$ transition. The cooler and repumper are each sent to their respective acousto-optic modulators (AOMs) to shift the cooler frequency $6\Gamma_0$ to the red of the (unshifted) $F=2\to F'=3$ transition and the repumper is set resonant with the (unshifted) $F=1\to F'=2$ transition. The pump-probe light is split from the cooler light and it passes through a pair of AOMs, used to tune the frequency to the in-trap resonance frequency of the $F=2, m_F=-2\to F'=3, m_{F'}=-3$ transition. This beam is then power-stabilized and split into two paths in free space, to form individual pump and probe beams. The pump is combined with the MOT beam path along $Z$ and the probe is fiber-coupled and sent through the pair of the lenses orthogonal to the FORT. The pump power is set via the power lock, while the probe power is adjusted independently with a variable ND filter, yielding probe counts rates at the detector ranging from \SIrange{0.5}{50}{\ctms}. The pump power was \SI{1.32}{\uW} and beam diameter of $\SI{2}{\mm}$, corresponding to an average pump-induced scattering of \SI{3.0}{\ctms}. The low level of pump-probe beams in the sequence (\autoref{fig:Sequence}) is equivalent to the pump power of \SI{55(2)}{\nW} before  the vacuum chamber, causing negligible photon scattering, and probe counts well below the dark counts of the detector. Photon counting is performed with near-infrared SPCMs (Excelitas, model SPCM-780-44-FC) with \SI{70}{\percent} quantum efficiency at \SI{780}{\nm} and dark count rate of \SI{40}{\cps}. Time tagging is performed by an Arty A7-35T FPGA. In the experimental sequence, we vary the magnetic field between $(B_X,B_Y,B_Z)=(0,0,182)\si{\milli\gauss}$ for trapping, $(B_X,B_Y,B_Z)=(0,0,0)\si{\milli\gauss}$ for cooling, and $(B_X,B_Y,B_Z)=(0,0,738)\si{\milli\gauss}$ for optical pumping and pump-probe exposure.} 

\begin{figure}
    \centering
    \includegraphics[width=0.85\columnwidth]{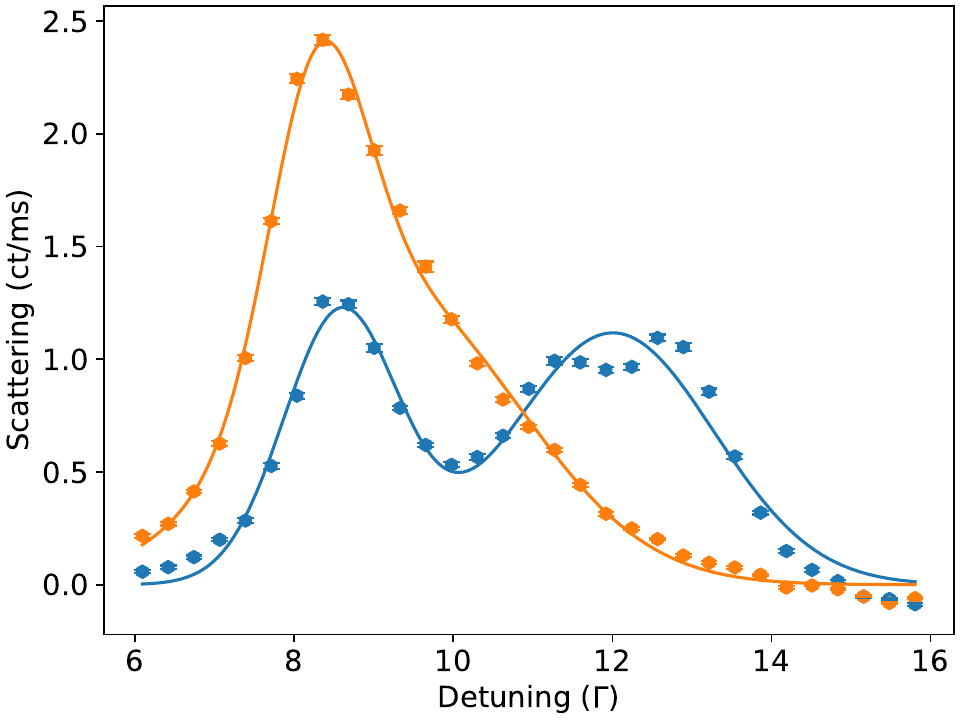}
    \caption{Pump-scattering spectra at zero magnetic field (blue) and at \SI{738}{\milli\gauss} (orange) along the $Z$ axis. Points show data and lines show fits with the sum of two Gaussians. Error bars show $\pm 1$ standard error of the mean.}
    \label{fig:opt_pumping}
\end{figure}

\PRLsection{Optical pumping} To achieve efficient population transfer to the stretched state while the magnetic sub-levels are far-detuned from each other due to the tensor light shift introduced by the FORT, we use ``ramp 1'' of  \autoref{fig:Sequence} to send \SI{55(2)}{\nW} of pump to the atom, detuned by $\approx12\Gamma_0$ to the blue of the natural transition. This increases the interaction between the pump and the magnetic sub-levels of $F'=3, m_{F'}\in[-2,2]$ which, for the FORT input power of $\SI{12}{\mW}$, are $\approx4\Gamma_0$ further detuned from the natural transition than are the $F'=3, m_{F'}\in\{-3,3\}$ levels. A magnetic field bias along $Z$, the  pump propagation axis, boosts the stretched-state scattering and suppresses scattering from higher-frequency transitions (see  \autoref{fig:opt_pumping}).  

\PRLsection{Effect of path fluctuations} 
\label{sec:ModelWithFluctuations}
We start with \autoref{eq:number_of_detected_phtons_averaging} and write the parameters in terms of means and fluctuations $\Nscbar = \Nscbarbar + \delta \Nscbar$, $\Nprbar = \Nprbarbar + {\delta}\Nprbar$, and $\Cbar = \Cbarbar + \delta \Cbar$. We take the fluctuations to be independent, zero-mean normal deviates, with respective variances $\sigma_{\rm sc}^2$, $\sigma_{\rm pr}^2$ and $\sigma_{C}^2$.  The detected signal then becomes
\begin{align}
    \label{eq:model_with_fluc}
    \Ncollbar &= (\Nscbarbar+\delta \Nscbar) + (\Nprbarbar+\delta \Nprbar) - 2 \cos\pathphase \nonumber \\ 
    & \times \sqrt{(\Nscbarbar+\delta \Nscbar)(\Nprbarbar +\delta \Nprbar)} (\Cbarbar +\delta \Cbar).
\end{align}
Using $\sqrt{(A+\delta A)(B+\delta B)} \simeq \sqrt{AB} + \sqrt{B/A} \delta A /2 + \sqrt{A/B} \delta B /2$ and neglecting second-order terms in Eq. (\autoref{eq:model_with_fluc}) leads to
\begin{equation} 
    \label{eq:counts_with_noise}
    \Ncollbar \simeq \Nscbarbar + \Nprbarbar - 2\sqrt{\Nscbarbar \Nprbarbar} \Cbarbar \cos \pathphase + \etanoise,
\end{equation}
where to first order the parameter fluctuations contribute 
\begin{equation}
    \begin{aligned}
    \etanoise  &= \delta \Nscbar \left( 1 - \sqrt{\frac{\Nprbarbar}{\Nscbarbar}} \Cbarbar\cos \pathphase \right)  \\
        & + \delta \Nprbar \left( 1 - \sqrt{\frac{\Nscbarbar}{\Nprbarbar}} \Cbarbar\cos \pathphase \right)  \\ 
        & - 2\sqrt{\Nscbarbar\Nprbarbar} \delta \Cbarbar  \cos \pathphase.
    \end{aligned}
\end{equation}

The variance of $\etanoise$, calculated using $\langle \cos^2\pathphase \rangle = 1/2$ for a uniformly distributed $\pathphase$, is
\begin{equation}
    \begin{aligned}
    \label{eq:total_variance}
    \sigma_n^2 \equiv \mathrm{var}(\eta_{\rm noise}) &=  \left(1+\frac{\Nprbarbar}{2\Nscbarbar}\Cbarbar^2\right)\sigma_{\rm sc}^2   \\
    & +  \left( 1+\frac{\Nscbarbar}{2\Nprbarbar}\Cbarbar^2 \right)\sigma_{\rm pr}^2+ 2\Nscbarbar\Nprbarbar\sigma_{C}^2.
    \end{aligned}
\end{equation}
\autoref{fig:fitted_sigmas} shows that the fitted $\sigma_n$ depends on the ratio $\Npr/\Nsc$, but is at most weakly dependent on time bin even as the temperature increases by nearly an order of magnitude. Noting that $\Cbarbar$ goes to zero as temperature increases, we conclude the dominant contributions are from changes in $\Nscbar$ and $\Nprbar$, rather than $\Cbar$.

\PRLsection{Heating model}
Using the R$\&$R method, we measured the atom temperature before and after the pump-probe exposure, finding $\temperature_\mathrm{init}=\SI{18(2)}{\micro\K}$ and $\temperature_\mathrm{end}=\SI{133(14)}{\micro\K}$, respectively. The main source of heating in the system is resonant scattering of pump photons.  \autoref{fig:temp_mapping} shows the collected scattering during a \SI{1}{\ms} pump exposure. The scattering decreases with time as the atom heats, spends less time at high FORT intensities, and experiences weaker light shifts, thus detuning the atomic resonance relative to the pump and probe. 

We model the scattering rate as
\begin{equation}
\label{eq:scat_decay_exp}
\Nsc(t) = \Nsc(0) e^{-\xi t}
\end{equation}
with $\Nsc(0) = \SI{3.68}{\ctms}$ and $\xi = \SI{0.51}{\per\second}$ found by fitting. The temperature is then 
\begin{equation}
\temperature(t) = \temperature_\mathrm{init} + \alpha \int_0^t  \Nsc(t') dt'
\end{equation}
where $\alpha = \SI{40.18}{\micro\kelvin\per ct}$ is the conversion from scattering rate to heating rate, chosen to give $\temperature(\SI{1}{\ms}) = \temperature_\mathrm{end}$.  

\clearpage
\end{document}